\documentclass[prd,amsfonts,amsmath,reprint,nofootinbib,floatfix]{revtex4-1}

\usepackage{color}
\usepackage{graphicx} 
\usepackage{slashed}
\usepackage{epstopdf}

\usepackage{mathptmx}   

\newcommand {\be} {\begin{equation}} 
\newcommand {\ba}{\begin{eqnarray}} 
\newcommand {\ee} {\end{equation}} 
\newcommand{\ea} {\end{eqnarray}}

\renewcommand{\Im}{{\rm Im\,}} 

\renewcommand{\epsilon}{\varepsilon}

\begin{document}

\title{Electron Beam Single-Spin Asymmetries in the Resonance Region with Final Hadrons Observed
}

\author{Brandon Buncher}

\author{Carl E.~Carlson}

\affiliation{Department of Physics, College of William and Mary, Williamsburg, VA 23187, USA}

\date{\today}

\begin{abstract}
We study beam-transverse and normal single-spin asymmetry for electron-proton inelastic scattering in the resonance region for the case where a pion is observed in the final state.  There is no beam single-spin asymmetry if the observed particle has single-photon exchange as its only interaction, but this is naturally circumvented when one observes a strongly interacting final state particle.  The asymmetries are not large as there is an electron mass divided by momentum transfer suppression. We present results for the cases where the final electron and a pion are both observed, with asymmetries of order tens of ppm, and where only a final pion is observed, with asymmetries of order one ppm.
\end{abstract}

\maketitle


\section{Introduction}			\label{sec:intro}


Beam single-spin asymmetries in an electron-hadron scattering process occur when just the initial electron is polarized, and no final state polarizations are measured.  Single-spin asymmetries are zero in one-photon exchange QED, barring further interactions.  None-the-less, they have been observed, indicating that additional interactions do occur.  Understanding these asymmetries is important, in particular to properly use electron scattering data to obtain fundamental metrics of hadronic structure such as nucleon electric and magnetic form factors, or the neutral weak interaction coupling to the proton.

More particularly, this paper will discuss beam single-spin asymmetry (BSSA) for normal and sideways electron polarization, where we define ``asymmetry'' as preferential scattering in a direction correlated with the initial polarization.  ``Normal'' indicates that the polarization is perpendicular to the scattering plane, while ``sideways'' indicates that the polarization remains in the scattering plane and perpendicular to the incoming electron's momentum.  Longitudinal spin asymmetries require parity violation, and do not occur if only QED and strong interactions are considered.  These violations may occur when weak interactions are considered, such as in events in which a $Z$-boson is exchanged.  Events such as these have been often utilized in order to measure the strangeness form factors and weak charge of the proton, notably recently by the QWeak experiment (see~\cite{Androic:2013rhu} and further references contained there).  Normal and sideways BSSA's are not forbidden by parity in general, although the sideways BSSA is forbidden by parity if only one final state particle is observed~\cite{Pasquini:2004pv}.  The normal or sideways BSSA expressed in terms of helicity amplitudes show a proportionality to the imaginary part of the interference term.  For relative phases to occur between the helicity amplitudes, there must be an additional interaction beyond the one-photon exchange, and for an asymmetry to be visible, the observed final state particle must have participated in the additional interaction.

When the electron is the observed final state particle, the additional interaction is necessarily a second photon exchange.  There exists a history suggesting that two-photon exchange effects are responsible for the once mysterious contradiction between the proton electric form factor obtained from differential cross section experiments and that obtained from polarization 
experiments~\cite{Blunden:2003sp,Arrington:2011dn,Chen:2004tw,Afanasev:2005mp,Carlson:2007sp,Borisyuk:2006fh,Borisyuk:2006uq}. Some theoretical results on two-photon exchange in the resonance region have also been reported~\cite{Pasquini(2015)}.  The existence of two-photon exchange is confirmed by experiments finding target-normal single-spin asymmetry (also forbidden in one-photon exchange) in $e$\,$^3$He quasielastic scattering with the final electron observed~\cite{Zhang:2015kna},  by the measured differences in $e^+$$p$ and $e^-$$p$ elastic scattering~\cite{Rachek:2014fam,Adikaram:2014ykv}, and by observed beam-normal single-spin asymmetries for $e$$p$ scattering both elastically~\cite{Waidyawansa:2013yva,Waidyawansa:2013pxa} and in the resonance region~~\cite{Nuruzzaman:2015vba}.

Beyond elastic scattering, it is possible to independently observe particles other than or in addition to the electron.  Even in inelastic scattering, if only the electron is observed and weak interactions are not considered, BSSA's remain forbidden in one-photon exchange.  However, if the reaction considered is $ep \to e\pi N$, where $N$ is a nucleon, one may observe the pion (say) instead of or along with the electron.  The pion necessarily experiences strong final state interactions, and these interactions give different phases to the amplitudes for the two electron helicities, and lead to non-zero beam-normal spin asymmetries.

In this paper, we calculate first the beam-spin asymmetries for the case where the pion and electron are both observed, and and show results when the final hadronic state is in the region of the $\Delta(1232)$ resonance. Then we calculate also for the case where solely the pion is observed, in the resonance region more generally. The methods used in the calculation follow standard procedure (reviewed, for example, in~\cite{Drechsel:1992pn}), but the results for the normal and sideways BSSA's are new, to the best of our knowledge.  Observing both the pion and the electron leads to an easier calculation since one needs to do no phase space integrals, but is more troublesome experimentally, as binning in two sets of variables leads to a paucity of data per bin.  In contrast, the calculation with solely the pion observed requires phase space integrals, which become somewhat tedious due to the necessity of finding connections between many kinematic variables and the integration variables one chooses.  However, one now will have significantly more data per bin.  Additionally, and importantly, the result can be useful for estimating contamination due to pion mis-identification in electron-observed beam-spin asymmetry experiments.  

We do not calculate \textit{ab initio} the strong phases the amplitudes acquire from the final state interactions, but obtain them by writing the results in terms of multipole amplitudes, and taking the multipole amplitudes from the MAID analysis~\cite{Drechsel:2007if}.

In the next section, we show the calculations and selected results for beam-spin asymmetries for the case where both the pion and electron are observed, and in Sec.~\ref{sec:pionly} we do the same for the case in which only the pion is observed.
Concluding remarks are offered in Sec.~\ref{sec:end} and some details of the calculation are shown in an Appendix.


\section{Beam single-spin asymmetry with electron and pion observed}
\label{sec:two}

\begin{figure}[b]

\begin{center}
\includegraphics{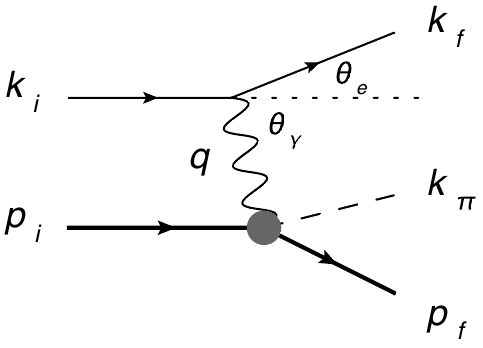}
\caption{Diagram of the reaction $e p \rightarrow e' N \pi$.  The electron interacts with the proton, exciting it to the $\Delta(1232)$ resonance, which decays into a nucleon and a pion.  
}
\label{fig:kin}
\end{center}
\end{figure}

To begin investigating the reaction $e p \rightarrow e' N \pi$ in the resonance region (Fig.~\ref{fig:kin}), we study the case in which both the outgoing electron and pion are observed.  This simplifies the calculation by removing the necessity of phase space integrals, and creates a reference to compare the next section's results to. To perform these calculations, the standard multipole amplitudes of pion electroproduction are utilized to determine the cross section of electron-proton scattering when pions are produced, and the cross sections are used to find the beam-normal and beam-sideways asymmetries.  The results are calculated numerically and plotted using Mathematica.  These calculations are performed using the one-photon exchange approximation.  Additionally, the electron mass has been neglected.   


As displayed in Fig.~\ref{fig:kin}, $k_i$ and $k_f$ refer to the incoming and outgoing electron's momenta, respectively, $q$ refers to the momentum transfer from the electron to the proton (with $Q^2 = -q^2$), $k_{\pi}$ refers to the pion's momentum, and $p_i$ and $p_f$ refer to the incoming and outgoing nucleons's momenta.  Further, $\theta_e$ refers to the outgoing electron's polar angle (its angle within the scattering plane), $\theta_{\gamma}$ refers to the angle between the photon and incoming electron, and the pion emerges at angles $\theta_\pi,\phi_\pi$, which may also be given in the $\pi$-final nucleon center-of-mass as $\theta_\pi^{\rm CM},\phi_\pi^{\rm CM}=\phi_\pi $.  In addition, several kinematic specifications are held fixed when performing calculations.  These are the incoming electron energy, $\epsilon_i$, the outgoing electron angle, $\theta_e$, and the final state hadron mass, $W$ or $m_X$.  In our numerical examples, matching current conditions in the QWeak experiment, we will take $\theta_e$ to be $7.9^\circ$ and $\epsilon_i$ to be 1.155 GeV.  Also, in this section, we will choose the final state hadron mass to be 1.232 GeV/$c^2$, corresponding to the mass of the $\Delta(1232)$ particle.

In order to calculate the beam-normal single-spin asymmetry (BNSSA, referring to the asymmetry generated by an electron polarized normally) and beam-sideways single-spin asymmetry (BSSSA, referring to the asymmetry generated by an electron polarized sideways) in which the electron and pion are observed, 
the differential cross section must first be calculated.  The full differential cross section for single-pion production in the one-photon exchange approximation is commonly given as (see for example~\cite{Drechsel:1992pn,Drechsel:2002ar}),
\begin{align}
\frac{d\sigma}{d^3 k_f \, d\Omega_\pi^{\rm CM}} = \epsilon_f^2 \Gamma \frac{d\sigma_{\rm v}}{d\Omega_\pi^{\rm CM}}  \,,
\end{align}
where
\begin{align}
    \Gamma = \frac{\alpha^2}{2\pi^2} \frac{\epsilon_f}{\epsilon_i} \frac{W^2-m_p^2}{2 m_p Q^2} \frac{1}{1-\epsilon} \,,
\end{align}
where $\epsilon$ is the photon polarization parameter (see the Appendix).  The cross section $d\sigma_v/d\Omega_\pi^{\rm CM}$ is for pion production from a virtual photon ($p_i \rightarrow p_f k_\pi$).

The BNSSA is,
\begin{equation}
B_n = \frac
{ \left. {d\sigma_{\rm v}}/{d\Omega^{\rm CM}_\pi} \right|_{s_n = 1} 
- \left. {d\sigma_{\rm v}}/{d\Omega^{\rm CM}_\pi} \right|_{s_n = -1} }
{ \left. {d\sigma_{\rm v}}/{d\Omega^{\rm CM}_\pi} \right|_{s_n = 1}  
+ \left. {d\sigma_{\rm v}}/{d\Omega^{\rm CM}_\pi} \right|_{s_n = -1}  },
\end{equation}
where $s_n = 1$ means the electron is polarized up (defined from $\vec k_i \times \vec k_f$), and down for $s_n = -1$.  The BSSSA can be similarly defined.  

Generally, if the beam electron is polarized perpendicular to its momentum direction, its spin will point in a direction given by an azimuthal angle $\phi_{Se}$, and the cross section may be written as
\begin{equation}
\label{eq:uns}
\frac{d \sigma_{\rm v}}{d \Omega_\pi^{\rm CM}} = \frac{d \sigma_{\rm v}^{\rm unpol}}{d \Omega_\pi^{\rm CM}} + \sin{\phi_{Se}} \frac{d \sigma_{\rm v}^{n}}{d \Omega_\pi^{\rm CM}} + \cos{\phi_{Se}} \frac{d \sigma_{\rm v}^{s}}{d \Omega_\pi^{\rm CM}}.
\end{equation}
The asymmetries become,
\begin{align}
B_n &=\frac{d \sigma_{\rm v}^{n} / d \Omega_\pi^{\rm CM}}{d \sigma_{\rm v}^{\rm unpol} / d \Omega_\pi^{\rm CM}},
    \nonumber\\
B_s &=\frac{d \sigma_{\rm v}^{s} / d \Omega_\pi^{\rm CM}}{d \sigma_{\rm v}^{\rm unpol} / d \Omega_\pi^{\rm CM}},
\end{align}

Formulas for the unpolarized cross sections, which can be located in many places (among them~\cite{Drechsel:1992pn} or~\cite{Haidan:1979yqa}), are
\begin{align}
\frac{d\sigma_{\rm v}^{\rm unpol}}{d\Omega^{\rm CM}_\pi}
	&=  A + \epsilon B + \epsilon C \sin^2\theta_\pi^{\rm CM} \cos 2\phi_\pi^{\rm CM} 
	    \nonumber\\
&\quad	+ \sqrt{2\epsilon (1+\epsilon) } D \sin\theta_\pi^{\rm CM} \cos\phi_\pi^{\rm CM} \nonumber\\
&= \frac{d\sigma_{\rm T}}{d\Omega^{\rm CM}_\pi} 
    + \epsilon \frac{Q^2}{\omega_\gamma^2} \frac{d\sigma_{\rm L}}{d\Omega^{\rm CM}_\pi}
    + \epsilon \frac{d\sigma_{\rm TT}}{d\Omega^{\rm CM}_\pi} \cos 2\phi_\pi^{\rm CM} \nonumber\\
&+    \sqrt{2\epsilon (1+\epsilon) } \frac{Q}{\omega_\gamma} 
          \frac{d\sigma_{\rm TL}}{d\Omega^{\rm CM}_\pi} \cos\phi_\pi^{\rm CM}   .
\end{align}
The cross sections with transverse beam polarizations are,
\begin{align}
\frac{d\sigma_{\rm v}^{n}}{d\Omega^{\rm CM}_\pi}
	&=   \frac{2m_e}{Q}  (1-\epsilon) D' \sin\theta_\pi^{\rm CM} \cos\phi_\pi^{\rm CM}   \,,	
	\nonumber\\
\frac{d\sigma_{\rm v}^{s}}{d\Omega^{\rm CM}_\pi}
	&=   \frac{2m_e}{Q}  (1-\epsilon) E' \cos\theta_\gamma \sin\theta_\pi^{\rm CM} 
		\sin\phi_\pi^{\rm CM} 	\,.
\end{align}
The $A$, $B$, $C$, and $D$ are the notation of Haidan~\cite{Haidan:1979yqa} and are provided in terms of multipole amplitudes, and equivalent expressions using $d\sigma_{IJ}/d\Omega^{\rm CM}_\pi$ can be found in Drechsel and Tiator~\cite{Drechsel:1992pn}.  $D'$ and $E'$, which were not found in previous studies, are
\begin{align}
\label{eq:E}
E' &= E'_0 + E'_1 \cos\theta_\pi^{\rm CM}	\,,	\\
	&E'_0 = \frac{|\vec k_\pi^{\rm CM}|}{k_\gamma^{\rm CM}} \frac{Q}{\omega_\gamma} \, \Im 
	\big\{ L_{0+}^*  (3E_{1+} - M_{1+} + M_{1-}) 
	                \nonumber\\
&\hskip 7 em	- (2L_{1+} - L_{1-})^* E_{0+} \big\}
		\,,	\nonumber\\
	&E'_1 = \frac{|\vec k_\pi^{\rm CM}|}{k_\gamma^{\rm CM}} \frac{Q}{\omega_\gamma} \,6\,\Im
	\left[ L_{1+}^*  ( E_{1+} - M_{1+} + M_{1-}) + L_{1-}^* E_{1+}
	\right]	\,,     \nonumber\\
D' &= - E' \,.
\end{align}
One can check that $D'$ is identical to $D$ in Haidan~\cite{Haidan:1979yqa} except for replacing ``Re'' with ``Im.''  Here, $\vec k_\pi^{\rm CM}$ refers to the pion momentum vector, and $\omega_\gamma$ to the photon lab energy.  Additionally, the $\mathcal M_{l\pm}$ refer to multipole amplitudes~\cite{Drechsel:1992pn} for $\gamma p \to \pi N$.  (As a reminder, $l$ is the orbital angular momentum of the pion, in the $\pi$-$N$ CM; ``$l_\pm$'' refers to the total angular momentum $j=l\pm 1/2$; $E_{l\pm}$ and $M_{l\pm}$ describe transversely polarized photons and $L_{l\pm}$ describes a longitudinally polarized photon; the angular momentum of the photon is $L$, so that $j=L\pm 1/2$, with $L=l$ for $M_{l\pm}$ and $|L-l|=1$ for $E_{l\pm}$ and $L_{l\pm}$.)


\begin{figure}[t]
\includegraphics[width=1\linewidth]{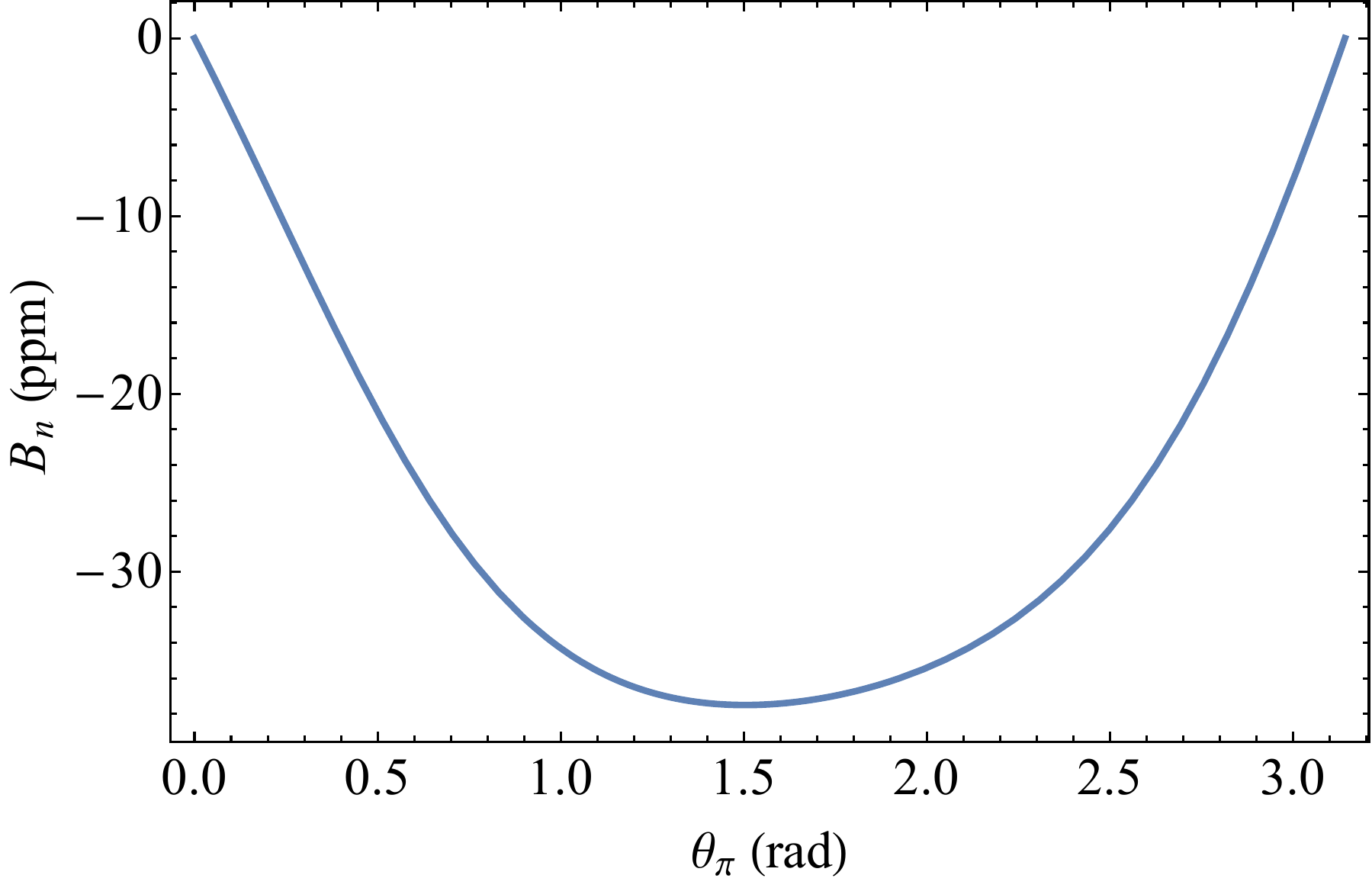}
\centerline{\quad \normalsize (a)}
\vskip 3 mm

\includegraphics[width=1\linewidth]{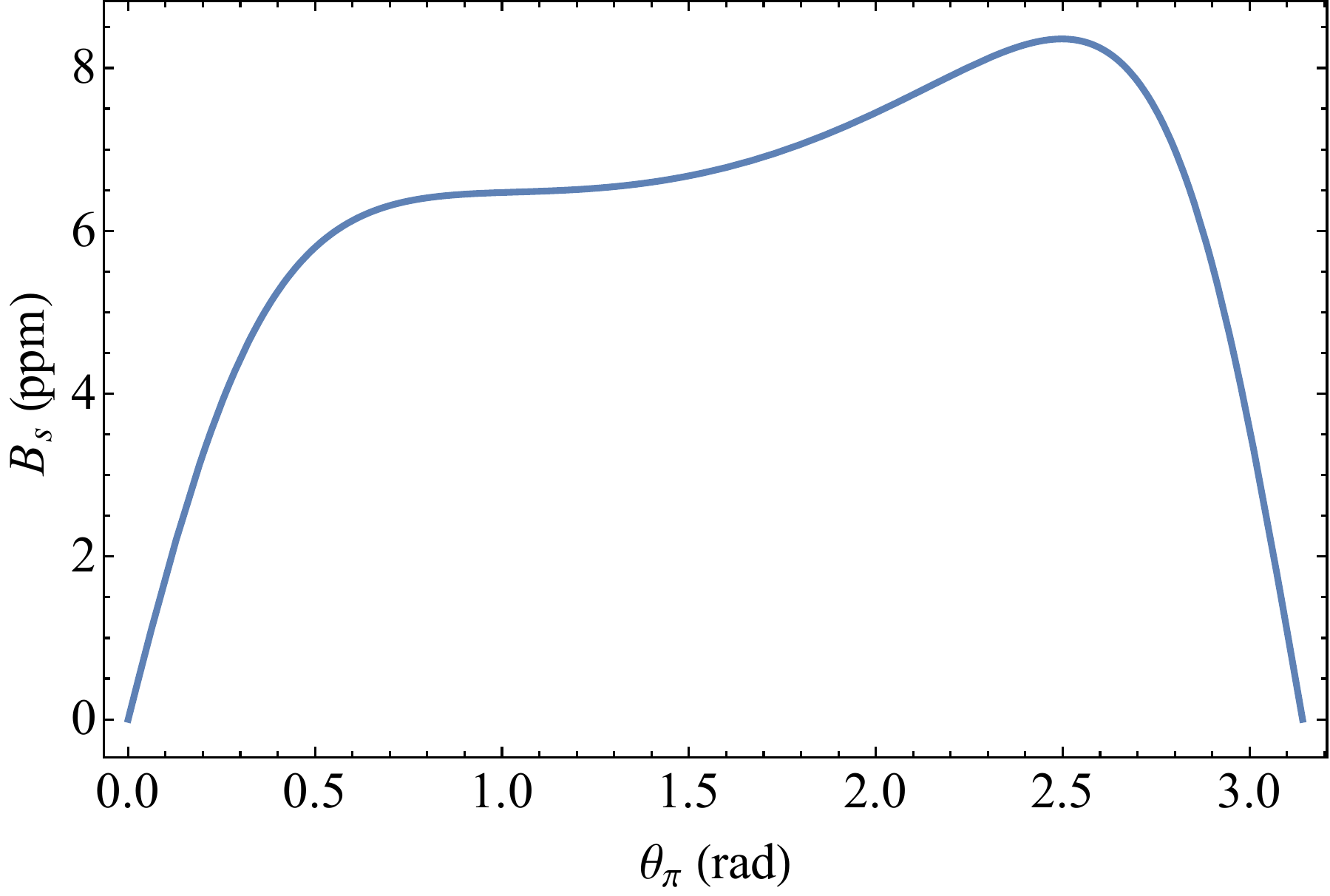}
\centerline{\normalsize (b)}

\caption{(a) Plot of BNSSA for $ep\to e \pi N$ with $e$ and $\pi$ observed and both $\pi^+$ and $\pi^0$ included.       $B_n$ is in parts per million and is plotted \textit{vs.}~the pion polar angle when the pion azimuthal angle is fixed at 0 radians.  (b) Similar plot of BSSSA $B_s$ in parts per million \textit{vs.}~the pion polar angle when the pion azimuthal angle is fixed at $\pi/2$ radians.  Both plots are for $\epsilon_i = 1.155$ GeV, $\theta_e = 7.9^\circ$, and $W = 1.232$ GeV.}

\label{fig:Ng2}

\end{figure}


We present numerical results for these asymmetries calculated using the values of $\epsilon_i$, $\theta_e$, and $W$ mentioned earlier ($1.155$  GeV, $7.9^\circ$, and $1.232$ GeV, respectively), and then plotted versus the pion polar angle $\theta_\pi^{\rm CM}$, for selected values of the pion azimuthal angle $\phi_\pi$.  Two plots are displayed in Fig.~\ref{fig:Ng2}. The plots are for the sum of $\pi^+$ and $\pi^0$ observed.  Fig.~\ref{fig:Ng2}(a) displays a BNSSA $B_n$ corresponding to a pion azimuthal angle of $0$ radians, and is approximately sinusoidal with a maximum magnitude at approximately $\theta_\pi=\frac{\pi}{2}$ rad of approximately 37 ppm.   Fig.~\ref{fig:Ng2}(b) corresponds to a BSSSA $B_s$.  This asymmetry is zero if the the pion momentum remains in the electron scattering plane as shown here for a pion azimuthal angle of $\pi/2$ radians.   These asymmetries are large enough in magnitude to be detected in modern experiments with sufficient statistics.


\section{Beam single-spin asymmetry with only pion observed}			\label{sec:pionly}


For inelastic $ep$ scattering with only a pion observed in the final state, $e+p \to \pi + X$,  spin dependent asymmetries may arise using the one-photon exchange approximation.  In order for this asymmetry to exist, it is required that a final state interaction occurs.  Since at least one hadron must exist among the particles within ``$X$,'' the pion will necessarily interact in the final state.  We shall not calculate this interaction \textit{ab initio}, but obtain the amplitudes---the phase being particularly crucial here---from the MAID data fit~\cite{Drechsel:2007if}.

Since the $\pi$ is observed, the effective mass of $X$ is known; however, the distribution of momenta within $X$ is unknown.  The $X$ must contain at least one electron and one nucelon, and we shall therefore concentrate on kinematic regions where the electron and nucleon are the sole components of $X$. 

The asymmetry will be defined much as before.  We will quote our results in the lab frame, with the electron and pion defining the $x$-$z$ plane.  The incoming electron's momentum defines the positive $z$-axis, and the transverse component of the scattered pion's three-momentum defines the positive $x$-direction.  The external variables will be the energy of the incoming electron $\epsilon_i$, the energy $\omega_\pi$ (or magnitude of the three-momentum $k_\pi$) of the outgoing pion, and the scattering angle of the outgoing pion $\theta_\pi$, plus the polarization direction of the beam electron.  In this paper, we assume the latter to be transverse, and define its direction by an azimuthal angle $\phi_S$. The beam single-spin asymmetry is
\begin{align}
B(\phi_S) = \frac
{ \sigma(\phi_S) - \sigma(\phi_S+180^\circ) }
{ \sigma(\phi_S) + \sigma(\phi_S+180^\circ) }  ,
\end{align}
where $\sigma$ is a differential cross section, and that both $\sigma$ and $B$ depend also on $\epsilon_i$, $k_\pi$, and $\theta_\pi$ is to be noted (though not shown above).  In this notation, the normal and sideways beam single-spin asymmetries are $B_n = B(90^\circ)$ and $B_s = B(0^\circ)$, respectively.

We will outline our calculation, providing some details in the Appendix, and subsequently discuss some numerical results for kinematics relevant to Jefferson Lab, especially the QWeak experiment.

The entire process is $e p\to e\pi  N$, and with the pion momentum specified, along with energy-momentum conservation, two variables must be integrated over.  We choose these to be the direction angles of the outgoing electron.  These integrations are performed in the $e$-$N$ center-of-mass (CM) frame, where no kinematic constraints affect the integration range.  Another frame of interest is the $\pi$-$N$ center-of-mass frame, as  the $\gamma^* p \to \pi N$ scattering amplitudes obtained from are MAID~\cite{Drechsel:2007if} are naturally found in this frame.  The experiments are performed in the lab frame, so the incoming electron and outgoing pion momenta are specified in this frame.  A large portion of the effort and apparent complexity of the present calculation lies in transforming variables among these three frames.

The differential cross section can be found following procedures reported in, among others, the review~\cite{Drechsel:1992pn}. When only the pion is observed in the final state, the relevant result can be expressed as
\begin{align}				
\frac{d\sigma}{d^3k_\pi} = \frac{\alpha}{2\pi^2} \frac{\epsilon_f^*}{ \epsilon_i \omega_\pi m_p m_X }
	\int d\Omega_e^* \frac{W^2}{Q^2 (1-\epsilon) } 
		\frac{k_\gamma^{\rm CM}}{k^{\rm CM}_\pi} 
			\frac{d\sigma_{\rm v}}{d\Omega^{\rm CM}_\pi} .
\label{eq:xsctn}
\end{align}

The superscript ``$*$'' denotes quantities in the final $e$-$N$ CM, and ``CM'' denotes specifically quantities in the $\pi$-$N$ CM, and quantities without either of these superscripts are in the lab frame or are Lorentz covariants.  As examples, in the above equation, $\epsilon_f^*$ is the final electron energy in the $e$-$N$ CM (the energy of the electron in this frame can be inferred after observing the pion, although its direction cannot), and $\epsilon_i$ is the incoming electron energy in the lab.  Other quantities are defined in Sec.~\ref{sec:two}.  Further, $m_X$ is the effective mass of $X$, the energy of the $e$-$N$ system in its own CM, and $W$ is the energy of the $\pi$-$N$ pair in its own CM.  The cross sections $d\sigma_v/d\Omega_\pi^{\rm CM}$ are provided in~\cite{Drechsel:1992pn} for the unpolarized case and in Sec.~\ref{sec:two} of the present paper when the electron is polarized transversely.

We split the cross section $d\sigma_v/d\Omega_\pi^{\rm CM}$ into unpolarized, normal polarized, and sideways polarized contributions as in Eq.~(\ref{eq:uns}).
The (non-longitudinal) beam single-spin asymmetry is defined as
\begin{align}
B(\phi_S) = \frac{
    \int d\Omega_e^* \frac{W^2}{Q^2 (1-\epsilon) } 
		\frac{k_\gamma^{\rm CM}}{k^{\rm CM}_\pi}
	\left[  \sin{\phi_{Se}} 
	\frac{d \sigma_{\rm v}^{\rm n}}{d \Omega_\pi^{\rm CM}} + \cos{\phi_{Se}} 
	\frac{d \sigma_{\rm v}^{\rm s}}
	    {d \Omega_\pi^{\rm CM}}  
	\right]
    }{
    \int d\Omega_e^* \frac{W^2}{Q^2 (1-\epsilon) } 
		\frac{k_\gamma^{\rm CM}}{k^{CM}_\pi}
	\frac{d \sigma_{\rm v}^{\rm unpol}}
	    {d \Omega_\pi^{CM}}
    }.
\label{eq:ratio}
\end{align}
The azimuthal angle $\phi_{Se}$ is in a lab frame where the outgoing electron defines the scattering plane.  It is hence different from $\phi_S$ defined with respect to a plane fixed by the pion momentum.   Angle $\phi_S$ is fixed by experimental conditions; however, $\phi_{Se}$ varies with the integration variables $\theta_e^*$ and $\phi_e^*$, and this relationship must be worked out. Expressions for these connects can be found in the Appendix.

\begin{figure}[ht]


\includegraphics[width= 8.3 cm]{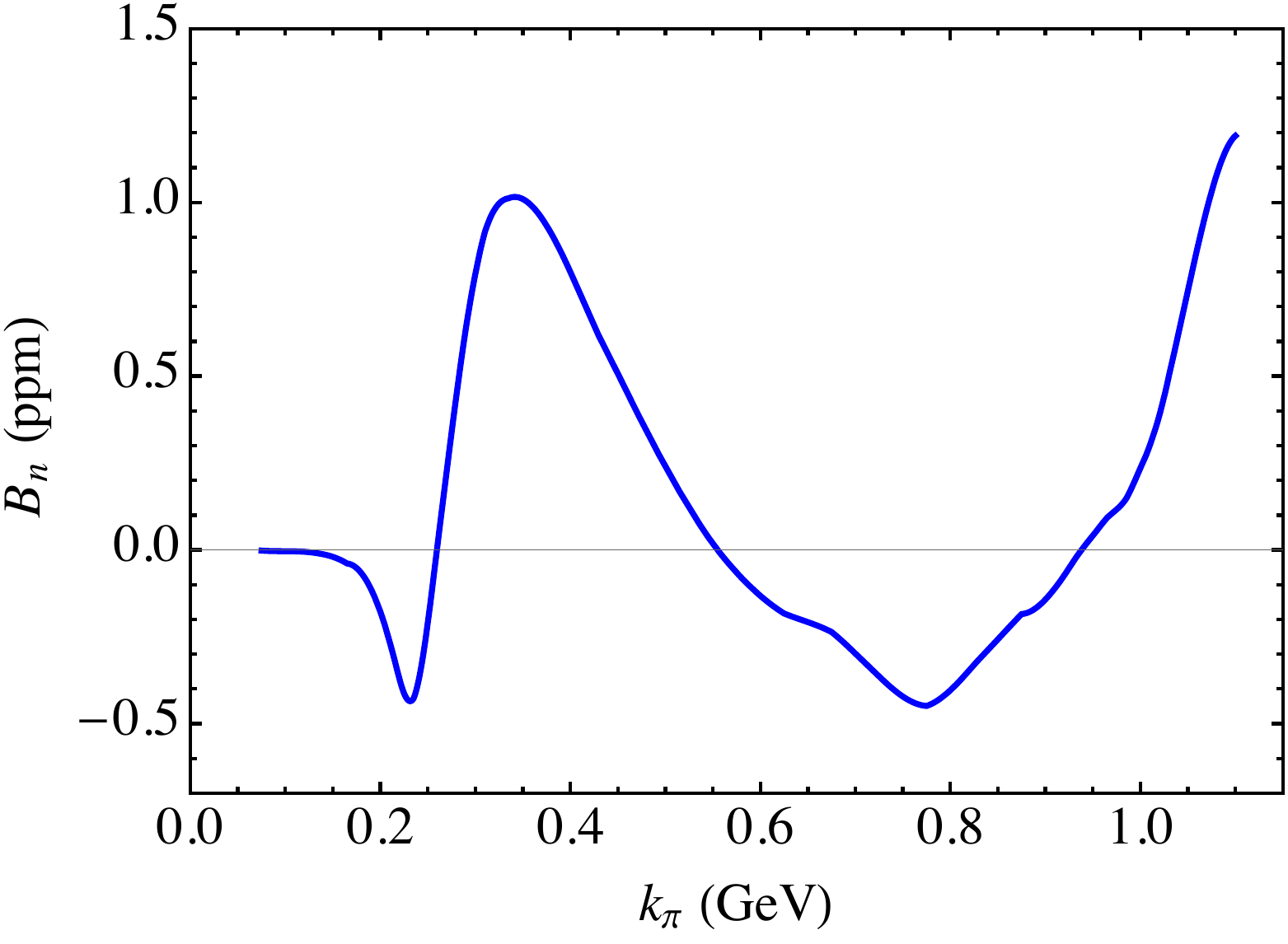}

\caption{
Beam-normal spin asymmetry for $e + p \to \pi^+ + X$. The plot is for incoming electron beam energy $\epsilon_i = 1.155$ GeV and for outgoing lab pion angle $7.9^\circ$. 
}
\label{fig:bnsa_pion}
\end{figure}

With only one particle observed in the final state, parity and rotation invariance suffice to prove that the sideways asymmetry $B_s = B(0^\circ)$ is zero.  We show the calculated beam-normal asymmetry in Fig.~\ref{fig:bnsa_pion} for the case that the pion is a $\pi^+$.  For definiteness, we examine kinematics such as those of the QWeak experiment, where the incoming electron energy is $\epsilon_i = 1.155$ GeV and the outgoing particle angle centered at about $7.9^\circ$.


The abscissa provides a range of pion momenta.  No difficulty arises when $k_\pi = 0$; rather, this merely describes the upper limit upon $m_X$,
\begin{align}
m_X \le (m_p-m_\pi)^2+ 2 \epsilon_i (m_p -m_\pi),
\end{align}
which is approximately $1.58$ GeV for the kinematics we use.  The upper limit of $k_\pi$ can be determined.  For sufficient energy, the pion mass can be neglected, yielding the approximation,
\begin{align}
k_\pi \le \frac{\epsilon_i}{1 + 2 (\epsilon_i/m_p) \sin^2 (\theta_\pi/2)}     ,
\end{align}
which numerically evaluates to $1.142$ GeV for present circumstances.  (Retaining a non-zero pion mass and using $m_{\pi^+}$ yields $k_\pi \le 1.132$ GeV.)

\begin{figure}[tb]

\includegraphics[width= 8.3 cm]{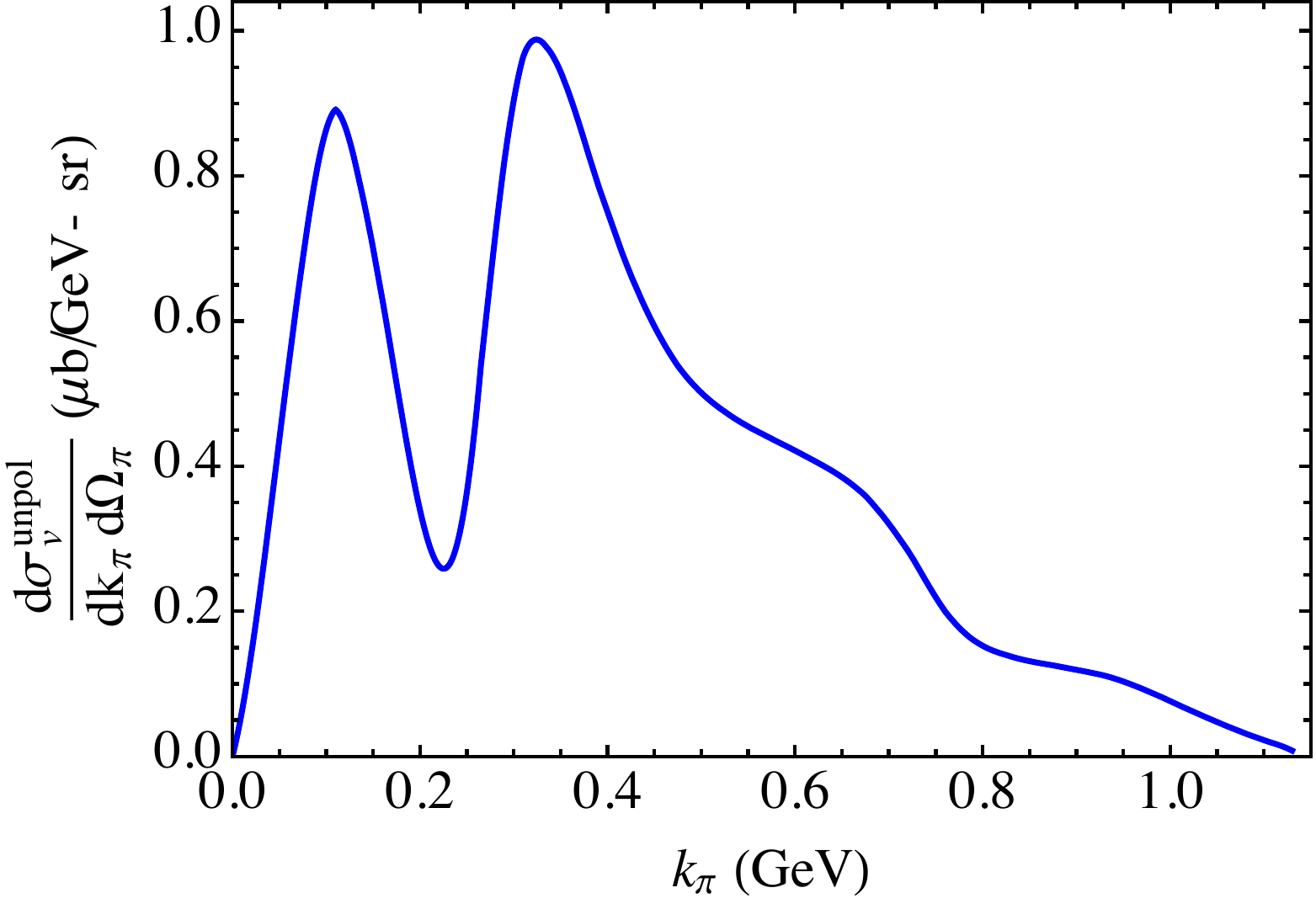}

\caption{
The unpolarized differential cross section for $\gamma^* + p \to \pi^+ + X$, with incoming electron beam energy $\epsilon_i = 1.155$ GeV and for outgoing lab pion angle $7.9^\circ$. 
}
\label{fig:xsctn}
\end{figure}

The unpolarized cross section for the same kinematics is displayed in Fig.~\ref{fig:xsctn}.  The first peak in this figure is due to kinematic rise and fall from threshold, and near the second peak the average of the hadronic mass $W$ is approximately that of the $\Delta(1232)$ resonance.  Similarly, for the shoulder visible from $k_\pi$ of approximately 0.6 to 0.75 GeV, the average of $W$ spans the range from the Roper$(1440)$ resonance to the $S_{11}(1520)$-$D_{13}(1535)$ resonance region.


\section{Closing Remarks}			\label{sec:end}


We have calculated the beam-spin asymmetries in the resonance region of the reaction $ep\rightarrow e'N\pi$.   

First, these calculations were performed for experiments in which both the electron and pion were observed. An asymmetry exists even using the one-photon exchange approximation because there are strong final state interactions involving a particle that is observed.  These calculations obtained their strong interaction information from the standard multipole amplitudes of pion electroproduction; the multipole amplitude formalism is reviewed in~\cite{Drechsel:1992pn} and numerical results for the amplitudes were obtained from the MAID parametrization~\cite{Drechsel:2007if}.  To our knowledge, the application of the formalism to the beam-normal and sideways single-spin asymmetries, which are suppressed by a factor of the electron mass over the momentum transfer, is new here.   The numerical results we presented were performed using conditions that follow the QWeak experiment: the final electron angle, $\theta_e$, was fixed at $7.9^\circ$, and the incoming electron energy, $\epsilon_i$, was fixed at 1.155 GeV. With two particles observed, we could fix the final state hadronic mass, and we fixed it to the $\Delta(1232)$ region.  The beam-normal asymmetry reached a peak magnitude of approximately 37 ppm, while the beam-sideways asymmetry attained a maximum value of approximately 16 ppm.  The magnitude of these asymmetries indicate that it would be observable, assuming appropriate instrumentation is utilized.  

Next, the same calculations were performed, again using the one-photon exchange approximation, for the case in which only the pion was observed. In this case we cannot specify the final hadronic mass, but can give averages for a given observed pion momentum. The numerical results we presented as examples of output used a pion angle the same as the outgoing electron angle above, and the same beam energy. For these kinematics, our maximum hadronic mass was 1.58 GeV, safely in the validity region of the MAID amplitudes.  Using these amplitudes, it was found that the asymmetry achieved a maximum magnitude on the order of 1 ppm.

Although our numerical examples used the QWeak kinematics, the outcome can be utilized for a variety of experiments.  One useful result resides in the magnitude of asymmetries observed.  In an electron-proton collision experiment, difficulty sometimes arises in the differentiation between an electron event and a pion event.  A large asymmetry inherently produced in the resonance region may produce a confounding variable in these experiments, which often utilize scattering asymmetries to study Weak interactions or search for beyond the standard model phenomena.  However, the asymmetries found in this study are very small, indicating that the beam-spin asymmetry effects will be negligible in most experiments.  This improves the security of many scattering experiments; however, experiments measuring asymmetries on the order of 10 ppm may need to correct for these effects.

Through this study, it was demonstrated that an asymmetry can be calculated from the one-photon exchange approximation.  Future studies could be devoted to utilizing this approximation in other scattering asymmetries.  In addition, the two-photon exchange approximation could be performed on this system to yield a more accurate description of beam-spin asymmetries when only the final electron is observed.


\appendix			



\section{Kinematic relations}			\label{sec:appendix}

\begin{figure}[ht]

\includegraphics[width= 8.3 cm]{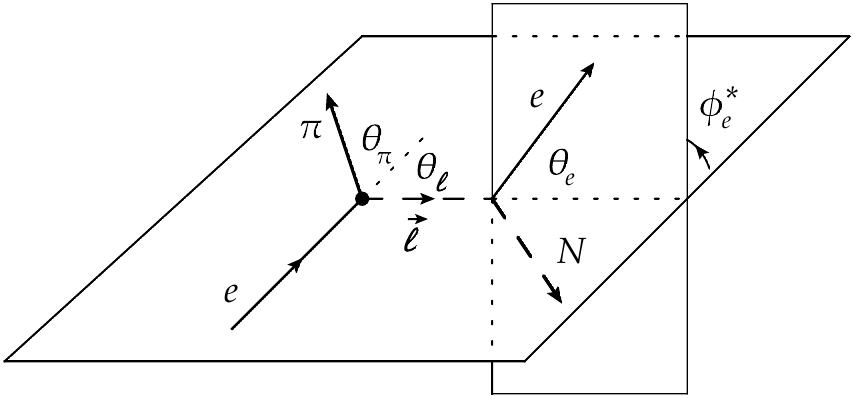}

\caption{The process $ep\to e \pi N$.  The incoming electron and observed outgoing pion define the scattering plane.  The final electron and final nucleon momenta, which are to be integrated over, lie in another plane rotated by an azimuthal angle $\phi_e = \phi_e^*$ to the scattering plane.  The dashed line labeled with momentum $\vec\ell$ solely indicates a momentum transfer (as opposed to a real or virtual particle.) }
\label{fig:kinematics}
\end{figure}

The process $e(k_i)+p(p_i)\to e(k_f)+\pi(k_\pi) + N(p_f)$ is diagrammed in Fig.~\ref{fig:kinematics}.  We consider the case where only the pion is observed, and the final state $X$ is $X = e + N = e_f +N_f$. Fixing the three-momentum of the pion, and considering energy-momentum conservation, there remains a two-dimensional integral over two unfixed final electron and nucleon momentum variables. By choice, the two variables are the polar and azimuthal angles of the electron (or nucleon) in the $e+N$ rest frame.  Choosing the angles in this frame allows the integrations to run over the full solid angle range, with no constraints from, for example, energy conservation.


We calculate a differential cross section with the initial momenta and pion momentum fixed.  In the lab frame where the initial electron momentum and final pion momentum define the $z$-axis and $x$-$z$ plane, the fixed momenta are given by $\epsilon_i$, $k_\pi$, and $\theta_\pi$.

The mass of the final $X=e+N$ state is obtained from
\begin{align}
m_X^2 &= (k_f + p_f)^2 = (p_i+k_i-k_\pi)^2 
        \nonumber\\
&= m_p^2 + m_\pi^2 + m_e^2 +2 m_p (\epsilon_i - \omega_\pi) - 2 \epsilon_i(\omega_\pi - k_\pi \cos\theta_\pi) \,,
\end{align}
from which the energies of the final electron and nucleon are given in the $e_f+N_f$ rest frame by
\begin{align}
\epsilon_f^* &= \frac{m_X^2 + m_e^2 - m_N^2}{2 m_X}\,,
\nonumber\\
E_f^* &= \frac{m_X^2 - m_e^2 + m_N^2}{2 m_X}    \,.
\end{align}

The connection between the lab frame and the $e_f$-$N_f$ CM is given by a Lorentz transformation along the direction given by $\vec \ell$, where
\begin{align}
\vec \ell = \vec k_f + \vec p_f 
= \vec k_i - \vec k_\pi .
\end{align}
The Lorentz transformation parameters are
\begin{align}
\gamma &= \frac{m_p+\epsilon_i - \omega_\pi}{m_X}\,, \quad
\beta = \sqrt{  1-\frac{1}{\gamma^2}  }	\,.
\end{align}

The integration variables are the direction angles $\{\theta_e^*,\phi_e^*\}$ of $\vec k_f^*$, the momentum of the final electron in the $e_f$-$N_f$ CM.  The polar angle is measured from the direction inherited from $\vec\ell$ and the azimuthal angle is relative to the original scattering plane. We must obtain quantities in the integrand of, for example, Eq.~(\ref{eq:xsctn}), in terms of the fixed momenta and the integration variables.  It is straightforward to boost $k_f$ back to the lab, to a lab frame where $\vec q$ is defining the $z$-direction, and then, after defining $\theta_\ell$ to be the angle between $\vec k_i$ and $\vec\ell$, to find
\begin{align}
Q^2 &= - (k_i-k_f)^2 			\nonumber\\
	&= -2m_e^2 + 2 \gamma \, \epsilon_f^*  ( \epsilon_i-\beta k_i \cos\theta_\ell)
	                            \nonumber\\
	& \hskip -1em + 2 \gamma k_f^* (\beta  \epsilon_i -k_i \cos\theta_\ell) \cos\theta_e^*
	- 2 k_i k_f^* \sin\theta_\ell \sin\theta_e^* \cos\phi_e^* 		\,,
							\nonumber\\
W^2 &= m_p^2 + 2m_p \left[ \epsilon_i -  \gamma (\epsilon_f^* +\beta k_f^* \cos\theta_e^*) \right]
	- Q^2 			.
\end{align}
We retain a finite, non-zero electron mass in $Q^2$ as configurations where the final electron was soft or collinear with the initial electron would lead to a singularity in the integrand assuming an electron mass of zero.  Retaining the electron mass ensures that $Q^2$ is never zero.   

The photon polarization parameter $\epsilon$ is provided in terms of the photon lab energy $\omega_\gamma$ and electron lab scattering angle $\theta_e$ (this angle relative to the initial electron direction rather than relative to $\vec q$),
\begin{align}
\frac{1}{\epsilon} = 1 + 
    2\left(1 + \frac{\omega_\gamma^2}{Q^2} \right)
    \tan^2\frac{\theta_e}{2}    \,.
\end{align}
The lab quantities are
\begin{align}
\omega_\gamma &= \epsilon_i - \epsilon_f
    = \epsilon_i - \gamma \epsilon^*_f
    (1 + \beta \cos\theta_e^* ) ,    \nonumber\\
\tan^2\frac{\theta_e}{2} &= 
    \frac{Q^2}{4\epsilon_i \epsilon_f - Q^2 }   \,.
\end{align}

Further, we require quantities in the pion-nucleon CM, where the $\gamma^* + p \to \pi + N$ multipole amplitude are obtained.  The factor $k_\gamma^{\rm CM}/k_\pi^{\rm CM}$ within the integrals in Eqs.~(\ref{eq:xsctn}) and (\ref{eq:ratio}) is not required, as it cancels a factor in the standard form of the expressions for the photoproduction cross sections (see, for example, Eq.~(\ref{eq:E})).   However, we do require an expression for the pion production angles in the $\pi$-$N_f$ CM.  We have derived these:
\begin{align}
\cos\theta_\pi^{\rm CM} &= \frac{ \omega_\pi^{\rm CM} \omega_\gamma^{\rm CM} - k_\pi \cdot q}
	{ k_\pi^{\rm CM} |\vec q ^{\rm \, CM} | }		\,,
\end{align}
where
\begin{align}
k_\pi \cdot q &= \omega_\pi \left( \epsilon_i - \epsilon_f^* \gamma (1+\beta\cos\theta_e^*) \right) \nonumber\\
&-	k_\pi \sin(\theta_\pi+\theta_\ell) 
	\left( \epsilon_i \sin\theta_\ell - \epsilon_f^* \sin\theta_e^* \cos\phi_e^* \right),	\nonumber\\
&-	k_\pi \cos(\theta_\pi+\theta_\ell)
	\left( \epsilon_i \cos\theta_\ell - \epsilon_f^* \gamma (\beta+\cos\theta_e^*) 	\right),
	                    \nonumber\\
\omega_\pi^{\rm CM} &= \frac{ W^2 + m_\pi^2 -m_p^2}{2W}	,
	\quad k_\pi^{CM} = \sqrt{ (\omega_\pi^{\rm CM})^2 - m_\pi^2}
	\nonumber\\
\omega^{\rm CM} &=  \frac{ W^2 - Q^2 -m_p^2}{2W}	 ,	
	\quad  |\vec q^{\rm \, CM}| = \sqrt{ (\omega^{\rm CM} )^2 + Q^2 }     ,
\end{align}
and
\begin{align}
\sin\phi_\pi^{CM} &= \frac{\epsilon_f^*}{\epsilon_f}
	\frac{ \sin\theta_\pi \sin\theta_e^* \sin\phi_e^* }{ \sin\theta_{\pi q} \sin\theta_e }	,   \nonumber\\
\cos\phi_\pi^{CM} &= \frac{ | \vec q \,|  \cos\theta_\pi 
	-  (\epsilon_i -  \epsilon_f \cos\theta_e )   \cos\theta_{\pi q}  }
	{ \epsilon_f  \sin\theta_{\pi q} \sin\theta_e  } .
\end{align}
Here $\theta_{\pi q}$ is the angle between $\vec k_\pi$ and $\vec q$ in a lab frame,
 \begin{align}
 \cos\theta_{\pi q} &= 
    \frac{ \omega_\pi  \omega_\gamma - k_\pi \cdot q}
	{  k_\pi  |\vec q  | }		\,.
 \end{align}

It remains to determine the polarization angles $\phi_{Se}$ in terms of $\phi_S$ and other quantities.  We have,
\begin{align}
\cos\phi_{Se} = &\big\{ \left[ \sin\theta_e^* \cos\phi_e^*   \cos\theta_\ell
	-  \gamma (\beta +  \cos\theta_e^* ) \sin\theta_\ell  \right] \cos\phi_S          \nonumber\\
&+   \sin\theta_e^* \sin\phi_e^* \sin\phi_S \big\} \big/ d_S
	\nonumber\\
\sin\phi_{Se} = &\big\{ \left[  \sin\theta_e^* \cos\phi_e^*   \cos\theta_\ell
	-  \gamma (\beta  +  \cos\theta_e^* ) \sin\theta_\ell  \right] \sin\phi_S  \nonumber\\
	&-  \left[\sin\theta_e^* \sin\phi_e^* \right] \cos\phi_S \big\} \big/ d_S      ,
\end{align}
where the denominator is
\begin{align}
d_S	= &\big\{ \left[  \sin\theta_e^* \cos\phi_e^*   \cos\theta_\ell
	-  \gamma (\beta  + \cos\theta_e^* ) \sin\theta_\ell  \right]^2       \nonumber\\
	&+ \left[  \sin\theta_e^* \sin\phi_e^* \right]^2
	\big\}^{1/2}		.
\end{align}


\begin{acknowledgments}

We thank the National Science Foundation for support under Grants PHY-1205905 and PHY-1516509.  We thank Mark Dalton, Mark Jones, and Lothar Tiator for help and comments.

\end{acknowledgments}

\bibliography{bssa2016}

\begin{thebibliography}{20}%
\makeatletter
\providecommand \@ifxundefined [1]{%
 \@ifx{#1\undefined}
}%
\providecommand \@ifnum [1]{%
 \ifnum #1\expandafter \@firstoftwo
 \else \expandafter \@secondoftwo
 \fi
}%
\providecommand \@ifx [1]{%
 \ifx #1\expandafter \@firstoftwo
 \else \expandafter \@secondoftwo
 \fi
}%
\providecommand \natexlab [1]{#1}%
\providecommand \enquote  [1]{``#1''}%
\providecommand \bibnamefont  [1]{#1}%
\providecommand \bibfnamefont [1]{#1}%
\providecommand \citenamefont [1]{#1}%
\providecommand \href@noop [0]{\@secondoftwo}%
\providecommand \href [0]{\begingroup \@sanitize@url \@href}%
\providecommand \@href[1]{\@@startlink{#1}\@@href}%
\providecommand \@@href[1]{\endgroup#1\@@endlink}%
\providecommand \@sanitize@url [0]{\catcode `\\12\catcode `\$12\catcode
  `\&12\catcode `\#12\catcode `\^12\catcode `\_12\catcode `\%12\relax}%
\providecommand \@@startlink[1]{}%
\providecommand \@@endlink[0]{}%
\providecommand \url  [0]{\begingroup\@sanitize@url \@url }%
\providecommand \@url [1]{\endgroup\@href {#1}{\urlprefix }}%
\providecommand \urlprefix  [0]{URL }%
\providecommand \Eprint [0]{\href }%
\providecommand \doibase [0]{http://dx.doi.org/}%
\providecommand \selectlanguage [0]{\@gobble}%
\providecommand \bibinfo  [0]{\@secondoftwo}%
\providecommand \bibfield  [0]{\@secondoftwo}%
\providecommand \translation [1]{[#1]}%
\providecommand \BibitemOpen [0]{}%
\providecommand \bibitemStop [0]{}%
\providecommand \bibitemNoStop [0]{.\EOS\space}%
\providecommand \EOS [0]{\spacefactor3000\relax}%
\providecommand \BibitemShut  [1]{\csname bibitem#1\endcsname}%
\let\auto@bib@innerbib\@empty
\bibitem [{\citenamefont {Androic}\ \emph {et~al.}(2013)\citenamefont {Androic}
  \emph {et~al.}}]{Androic:2013rhu}%
  \BibitemOpen
  \bibfield  {author} {\bibinfo {author} {\bibfnamefont {D.}~\bibnamefont
  {Androic}} \emph {et~al.} (\bibinfo {collaboration} {Qweak}),\ }\href
  {\doibase 10.1103/PhysRevLett.111.141803} {\bibfield  {journal} {\bibinfo
  {journal} {Phys. Rev. Lett.}\ }\textbf {\bibinfo {volume} {111}},\ \bibinfo
  {pages} {141803} (\bibinfo {year} {2013})},\ \Eprint
  {http://arxiv.org/abs/1307.5275} {arXiv:1307.5275 [nucl-ex]} \BibitemShut
  {NoStop}%
\bibitem [{\citenamefont {Pasquini}\ and\ \citenamefont
  {Vanderhaeghen}(2004)}]{Pasquini:2004pv}%
  \BibitemOpen
  \bibfield  {author} {\bibinfo {author} {\bibfnamefont {B.}~\bibnamefont
  {Pasquini}}\ and\ \bibinfo {author} {\bibfnamefont {M.}~\bibnamefont
  {Vanderhaeghen}},\ }\href {\doibase 10.1103/PhysRevC.70.045206} {\bibfield
  {journal} {\bibinfo  {journal} {Phys. Rev.}\ }\textbf {\bibinfo {volume}
  {C70}},\ \bibinfo {pages} {045206} (\bibinfo {year} {2004})},\ \Eprint
  {http://arxiv.org/abs/hep-ph/0405303} {arXiv:hep-ph/0405303 [hep-ph]}
  \BibitemShut {NoStop}%
\bibitem [{\citenamefont {Blunden}\ \emph {et~al.}(2003)\citenamefont
  {Blunden}, \citenamefont {Melnitchouk},\ and\ \citenamefont
  {Tjon}}]{Blunden:2003sp}%
  \BibitemOpen
  \bibfield  {author} {\bibinfo {author} {\bibfnamefont {P.~G.}\ \bibnamefont
  {Blunden}}, \bibinfo {author} {\bibfnamefont {W.}~\bibnamefont
  {Melnitchouk}}, \ and\ \bibinfo {author} {\bibfnamefont {J.~A.}\ \bibnamefont
  {Tjon}},\ }\href {\doibase 10.1103/PhysRevLett.91.142304} {\bibfield
  {journal} {\bibinfo  {journal} {Phys. Rev. Lett.}\ }\textbf {\bibinfo
  {volume} {91}},\ \bibinfo {pages} {142304} (\bibinfo {year} {2003})},\
  \Eprint {http://arxiv.org/abs/nucl-th/0306076} {arXiv:nucl-th/0306076
  [nucl-th]} \BibitemShut {NoStop}%
\bibitem [{\citenamefont {Arrington}\ \emph {et~al.}(2011)\citenamefont
  {Arrington}, \citenamefont {Blunden},\ and\ \citenamefont
  {Melnitchouk}}]{Arrington:2011dn}%
  \BibitemOpen
  \bibfield  {author} {\bibinfo {author} {\bibfnamefont {J.}~\bibnamefont
  {Arrington}}, \bibinfo {author} {\bibfnamefont {P.~G.}\ \bibnamefont
  {Blunden}}, \ and\ \bibinfo {author} {\bibfnamefont {W.}~\bibnamefont
  {Melnitchouk}},\ }\href {\doibase 10.1016/j.ppnp.2011.07.003} {\bibfield
  {journal} {\bibinfo  {journal} {Prog. Part. Nucl. Phys.}\ }\textbf {\bibinfo
  {volume} {66}},\ \bibinfo {pages} {782} (\bibinfo {year} {2011})},\ \Eprint
  {http://arxiv.org/abs/1105.0951} {arXiv:1105.0951 [nucl-th]} \BibitemShut
  {NoStop}%
\bibitem [{\citenamefont {Chen}\ \emph {et~al.}(2004)\citenamefont {Chen},
  \citenamefont {Afanasev}, \citenamefont {Brodsky}, \citenamefont {Carlson},\
  and\ \citenamefont {Vanderhaeghen}}]{Chen:2004tw}%
  \BibitemOpen
  \bibfield  {author} {\bibinfo {author} {\bibfnamefont {Y.~C.}\ \bibnamefont
  {Chen}}, \bibinfo {author} {\bibfnamefont {A.}~\bibnamefont {Afanasev}},
  \bibinfo {author} {\bibfnamefont {S.~J.}\ \bibnamefont {Brodsky}}, \bibinfo
  {author} {\bibfnamefont {C.~E.}\ \bibnamefont {Carlson}}, \ and\ \bibinfo
  {author} {\bibfnamefont {M.}~\bibnamefont {Vanderhaeghen}},\ }\href {\doibase
  10.1103/PhysRevLett.93.122301} {\bibfield  {journal} {\bibinfo  {journal}
  {Phys. Rev. Lett.}\ }\textbf {\bibinfo {volume} {93}},\ \bibinfo {pages}
  {122301} (\bibinfo {year} {2004})},\ \Eprint
  {http://arxiv.org/abs/hep-ph/0403058} {arXiv:hep-ph/0403058 [hep-ph]}
  \BibitemShut {NoStop}%
\bibitem [{\citenamefont {Afanasev}\ \emph {et~al.}(2005)\citenamefont
  {Afanasev}, \citenamefont {Brodsky}, \citenamefont {Carlson}, \citenamefont
  {Chen},\ and\ \citenamefont {Vanderhaeghen}}]{Afanasev:2005mp}%
  \BibitemOpen
  \bibfield  {author} {\bibinfo {author} {\bibfnamefont {A.~V.}\ \bibnamefont
  {Afanasev}}, \bibinfo {author} {\bibfnamefont {S.~J.}\ \bibnamefont
  {Brodsky}}, \bibinfo {author} {\bibfnamefont {C.~E.}\ \bibnamefont
  {Carlson}}, \bibinfo {author} {\bibfnamefont {Y.-C.}\ \bibnamefont {Chen}}, \
  and\ \bibinfo {author} {\bibfnamefont {M.}~\bibnamefont {Vanderhaeghen}},\
  }\href {\doibase 10.1103/PhysRevD.72.013008} {\bibfield  {journal} {\bibinfo
  {journal} {Phys. Rev.}\ }\textbf {\bibinfo {volume} {D72}},\ \bibinfo {pages}
  {013008} (\bibinfo {year} {2005})},\ \Eprint
  {http://arxiv.org/abs/hep-ph/0502013} {arXiv:hep-ph/0502013 [hep-ph]}
  \BibitemShut {NoStop}%
\bibitem [{\citenamefont {Carlson}\ and\ \citenamefont
  {Vanderhaeghen}(2007)}]{Carlson:2007sp}%
  \BibitemOpen
  \bibfield  {author} {\bibinfo {author} {\bibfnamefont {C.~E.}\ \bibnamefont
  {Carlson}}\ and\ \bibinfo {author} {\bibfnamefont {M.}~\bibnamefont
  {Vanderhaeghen}},\ }\href {\doibase 10.1146/annurev.nucl.57.090506.123116}
  {\bibfield  {journal} {\bibinfo  {journal} {Ann. Rev. Nucl. Part. Sci.}\
  }\textbf {\bibinfo {volume} {57}},\ \bibinfo {pages} {171} (\bibinfo {year}
  {2007})},\ \Eprint {http://arxiv.org/abs/hep-ph/0701272}
  {arXiv:hep-ph/0701272 [HEP-PH]} \BibitemShut {NoStop}%
\bibitem [{\citenamefont {Borisyuk}\ and\ \citenamefont
  {Kobushkin}(2006)}]{Borisyuk:2006fh}%
  \BibitemOpen
  \bibfield  {author} {\bibinfo {author} {\bibfnamefont {D.}~\bibnamefont
  {Borisyuk}}\ and\ \bibinfo {author} {\bibfnamefont {A.}~\bibnamefont
  {Kobushkin}},\ }\href {\doibase 10.1103/PhysRevC.74.065203} {\bibfield
  {journal} {\bibinfo  {journal} {Phys. Rev.}\ }\textbf {\bibinfo {volume}
  {C74}},\ \bibinfo {pages} {065203} (\bibinfo {year} {2006})},\ \Eprint
  {http://arxiv.org/abs/nucl-th/0606030} {arXiv:nucl-th/0606030 [nucl-th]}
  \BibitemShut {NoStop}%
\bibitem [{\citenamefont {Borisyuk}\ and\ \citenamefont
  {Kobushkin}(2007)}]{Borisyuk:2006uq}%
  \BibitemOpen
  \bibfield  {author} {\bibinfo {author} {\bibfnamefont {D.}~\bibnamefont
  {Borisyuk}}\ and\ \bibinfo {author} {\bibfnamefont {A.}~\bibnamefont
  {Kobushkin}},\ }\href {\doibase 10.1103/PhysRevC.75.038202} {\bibfield
  {journal} {\bibinfo  {journal} {Phys. Rev.}\ }\textbf {\bibinfo {volume}
  {C75}},\ \bibinfo {pages} {038202} (\bibinfo {year} {2007})},\ \Eprint
  {http://arxiv.org/abs/nucl-th/0612104} {arXiv:nucl-th/0612104 [nucl-th]}
  \BibitemShut {NoStop}%
\bibitem [{Pas()}]{Pasquini(2015)}%
  \BibitemOpen
  \href@noop {} {}\bibinfo {note} {B.~Pasquini, Talk at MAMI and Beyond
  Workshop, Mainz (2009) [http://conference.kph.uni-mainz.de/mamiandbeyond/]
  and private communication.}\BibitemShut {Stop}%
\bibitem [{\citenamefont {Zhang}\ \emph {et~al.}(2015)\citenamefont {Zhang}
  \emph {et~al.}}]{Zhang:2015kna}%
  \BibitemOpen
  \bibfield  {author} {\bibinfo {author} {\bibfnamefont {Y.~W.}\ \bibnamefont
  {Zhang}} \emph {et~al.},\ }\href {\doibase 10.1103/PhysRevLett.115.172502}
  {\bibfield  {journal} {\bibinfo  {journal} {Phys. Rev. Lett.}\ }\textbf
  {\bibinfo {volume} {115}},\ \bibinfo {pages} {172502} (\bibinfo {year}
  {2015})},\ \Eprint {http://arxiv.org/abs/1502.02636} {arXiv:1502.02636
  [nucl-ex]} \BibitemShut {NoStop}%
\bibitem [{\citenamefont {Rachek}\ \emph {et~al.}(2015)\citenamefont {Rachek}
  \emph {et~al.}}]{Rachek:2014fam}%
  \BibitemOpen
  \bibfield  {author} {\bibinfo {author} {\bibfnamefont {I.~A.}\ \bibnamefont
  {Rachek}} \emph {et~al.},\ }\href {\doibase 10.1103/PhysRevLett.114.062005}
  {\bibfield  {journal} {\bibinfo  {journal} {Phys. Rev. Lett.}\ }\textbf
  {\bibinfo {volume} {114}},\ \bibinfo {pages} {062005} (\bibinfo {year}
  {2015})},\ \Eprint {http://arxiv.org/abs/1411.7372} {arXiv:1411.7372
  [nucl-ex]} \BibitemShut {NoStop}%
\bibitem [{\citenamefont {Adikaram}\ \emph {et~al.}(2015)\citenamefont
  {Adikaram} \emph {et~al.}}]{Adikaram:2014ykv}%
  \BibitemOpen
  \bibfield  {author} {\bibinfo {author} {\bibfnamefont {D.}~\bibnamefont
  {Adikaram}} \emph {et~al.} (\bibinfo {collaboration} {CLAS}),\ }\href
  {\doibase 10.1103/PhysRevLett.114.062003} {\bibfield  {journal} {\bibinfo
  {journal} {Phys. Rev. Lett.}\ }\textbf {\bibinfo {volume} {114}},\ \bibinfo
  {pages} {062003} (\bibinfo {year} {2015})},\ \Eprint
  {http://arxiv.org/abs/1411.6908} {arXiv:1411.6908 [nucl-ex]} \BibitemShut
  {NoStop}%
\bibitem [{\citenamefont
  {Waidyawansa}(2013{\natexlab{a}})}]{Waidyawansa:2013yva}%
  \BibitemOpen
  \bibfield  {author} {\bibinfo {author} {\bibfnamefont {B.~P.}\ \bibnamefont
  {Waidyawansa}} (\bibinfo {collaboration} {QWeak}),\ }\bibfield  {booktitle}
  {\emph {\bibinfo {booktitle} {{Proceedings, 11th Conference on the
  Intersections of Particle and Nuclear Physics (CIPANP 2012)}}},\ }\href
  {\doibase 10.1063/1.4826847} {\bibfield  {journal} {\bibinfo  {journal} {AIP
  Conf. Proc.}\ }\textbf {\bibinfo {volume} {1560}},\ \bibinfo {pages} {583}
  (\bibinfo {year} {2013}{\natexlab{a}})}\BibitemShut {NoStop}%
\bibitem [{\citenamefont
  {Waidyawansa}(2013{\natexlab{b}})}]{Waidyawansa:2013pxa}%
  \BibitemOpen
  \bibfield  {author} {\bibinfo {author} {\bibfnamefont {D.~B.~P.}\
  \bibnamefont {Waidyawansa}},\ }\emph {\bibinfo {title} {{A 3\% Measurement of
  the Beam Normal Single Spin Asymmetry in Forward Angle Elastic
  Electron-Proton Scattering using the Q$_{\rm{weak}}$ Setup}}},\ \href
  {http://edwards1.phy.ohiou.edu/~roche/group_page/documents/reports/Waidyawansa_Thesis_July2013.pdf}
  {Ph.D. thesis},\ \bibinfo  {school} {Ohio U., Athens} (\bibinfo {year}
  {2013}{\natexlab{b}})\BibitemShut {NoStop}%
\bibitem [{\citenamefont {Nuruzzaman}(2015)}]{Nuruzzaman:2015vba}%
  \BibitemOpen
  \bibfield  {author} {\bibinfo {author} {\bibnamefont {Nuruzzaman}} (\bibinfo
  {collaboration} {QWeak}),\ }in\ \href
  {http://inspirehep.net/record/1395987/files/arXiv:1510.00449.pdf} {\emph
  {\bibinfo {booktitle} {{12th Conference on the Intersections of Particle and
  Nuclear Physics (CIPANP 2015) Vail, Colorado, USA, May 19-24, 2015}}}}\
  (\bibinfo {year} {2015})\ \Eprint {http://arxiv.org/abs/1510.00449}
  {arXiv:1510.00449 [nucl-ex]} \BibitemShut {NoStop}%
\bibitem [{\citenamefont {Drechsel}\ and\ \citenamefont
  {Tiator}(1992)}]{Drechsel:1992pn}%
  \BibitemOpen
  \bibfield  {author} {\bibinfo {author} {\bibfnamefont {D.}~\bibnamefont
  {Drechsel}}\ and\ \bibinfo {author} {\bibfnamefont {L.}~\bibnamefont
  {Tiator}},\ }\href {\doibase 10.1088/0954-3899/18/3/004} {\bibfield
  {journal} {\bibinfo  {journal} {J. Phys.}\ }\textbf {\bibinfo {volume}
  {G18}},\ \bibinfo {pages} {449} (\bibinfo {year} {1992})}\BibitemShut
  {NoStop}%
\bibitem [{\citenamefont {Drechsel}\ \emph {et~al.}(2007)\citenamefont
  {Drechsel}, \citenamefont {Kamalov},\ and\ \citenamefont
  {Tiator}}]{Drechsel:2007if}%
  \BibitemOpen
  \bibfield  {author} {\bibinfo {author} {\bibfnamefont {D.}~\bibnamefont
  {Drechsel}}, \bibinfo {author} {\bibfnamefont {S.~S.}\ \bibnamefont
  {Kamalov}}, \ and\ \bibinfo {author} {\bibfnamefont {L.}~\bibnamefont
  {Tiator}},\ }\href {\doibase 10.1140/epja/i2007-10490-6} {\bibfield
  {journal} {\bibinfo  {journal} {Eur. Phys. J.}\ }\textbf {\bibinfo {volume}
  {A34}},\ \bibinfo {pages} {69} (\bibinfo {year} {2007})},\ \Eprint
  {http://arxiv.org/abs/0710.0306} {arXiv:0710.0306 [nucl-th]} \BibitemShut
  {NoStop}%
\bibitem [{\citenamefont {Drechsel}\ \emph {et~al.}(2003)\citenamefont
  {Drechsel}, \citenamefont {Pasquini},\ and\ \citenamefont
  {Vanderhaeghen}}]{Drechsel:2002ar}%
  \BibitemOpen
  \bibfield  {author} {\bibinfo {author} {\bibfnamefont {D.}~\bibnamefont
  {Drechsel}}, \bibinfo {author} {\bibfnamefont {B.}~\bibnamefont {Pasquini}},
  \ and\ \bibinfo {author} {\bibfnamefont {M.}~\bibnamefont {Vanderhaeghen}},\
  }\href {\doibase 10.1016/S0370-1573(02)00636-1} {\bibfield  {journal}
  {\bibinfo  {journal} {Phys. Rept.}\ }\textbf {\bibinfo {volume} {378}},\
  \bibinfo {pages} {99} (\bibinfo {year} {2003})},\ \Eprint
  {http://arxiv.org/abs/hep-ph/0212124} {arXiv:hep-ph/0212124 [hep-ph]}
  \BibitemShut {NoStop}%
\bibitem [{\citenamefont {Haidan}(1979)}]{Haidan:1979yqa}%
  \BibitemOpen
  \bibfield  {author} {\bibinfo {author} {\bibfnamefont {R.}~\bibnamefont
  {Haidan}},\ }\emph {\bibinfo {title} {{Elektroproduktion pseudoskalarer
  Mesonen im Resonanzgebiet bei grossen Impuls\"ubertr\"agen}}},\ \href
  {http://www-library.desy.de/cgi-bin/showprep.pl?DESY-F21-79-03} {Ph.D.
  thesis},\ \bibinfo  {school} {Hamburg U.} (\bibinfo {year}
  {1979})\BibitemShut {NoStop}%
\end{thebibliography}%

\end{document}